\documentclass[a4paper,11pt]{article}
\pdfoutput=1

\usepackage{jinstpub}

\usepackage{siunitx}
\usepackage{subcaption}
\usepackage{url}

\title{\boldmath Performance of the new RD51 VMM3a/SRS beam telescope---studying MPGDs simultaneously in energy, space and time at high rates}

\author[a,b,1]{L. Scharenberg,\note{Corresponding author.}}
\author[c]{J. Bortfeldt,}
\author[a]{F. Brunbauer,}
\author[b]{K. Desch,}
\author[a,d]{K. Fl{\"o}thner,}
\author[e]{F. Garcia,}
\author[a,f,g]{D. Janssens,}
\author[a,h]{M. Lisowska,}
\author[b,a]{H. Muller,}
\author[a]{E. Oliveri,}
\author[a,i]{G. Orlandini,}
\author[j,k,a]{D. Pfeiffer,}
\author[a]{L. Ropelewski,}
\author[j,a]{J. Samarati,}
\author[l,a]{D. Sorvisto,}
\author[a]{M. van Stenis,}
\author[a]{and R. Veenhof}

\affiliation[a]{European Organization for Nuclear Research (CERN), 1211 Geneva 23, Switzerland}
\affiliation[b]{Physikalisches Institut, University of Bonn, Nu{\ss{}}allee 12, 53115 Bonn, Germany}
\affiliation[c]{Department for Medical Physics, Ludwig Maximilian University of Munich, Am Coulombwall 1, 85748 Garching, Germany}
\affiliation[d]{Helmholtz-Institut f{\"u}r Strahlen- und Kernphysik, University of Bonn, Nu{\ss{}}allee 14-16, 53115 Bonn, Germany}
\affiliation[e]{Helsinki Institute of Physics (HIP), P.O. Box 64, FI-00014 University of Helsinki, Finland}
\affiliation[f]{Inter-University Institute For High Energies, Pleinlaan 2, 1050 Brussels, Belgium}
\affiliation[g]{Vrije Universiteit Brussel, Pleinlaan 2, 1050 Brussels, Belgium}
\affiliation[h]{Universit{\'e} Paris-Saclay, 3 rue Joliot Curie, 91190 Gif-sur-Yvette, France}
\affiliation[i]{Friedrich-Alexander-Universit{\"a}t Erlangen-N{\"u}rnberg, Schlo\ss{}platz 4, 91054 Erlangen, Germany}
\affiliation[j]{European Spallation Source ERIC (ESS), Box 176, SE-221 00 Lund, Sweden}
\affiliation[k]{Dipartimento di Fisica, University of Milano-Bicocca, Piazza della Scienza 3, 20126 Milan, Italy}
\affiliation[l]{Aalto University, P.O. Box 11000, FI-00076 Aalto, Finland}

\emailAdd{lucian.scharenberg@cern.ch}

\abstract{The RD51 collaboration maintains a common infrastructure at CERN for its R \& D activities, including two beam telescopes for test beam campaigns.
    Recently, one of the beam telescopes has been equipped and commissioned with new multi-channel and charge-sensitive front-end electronics based on the ATLAS/BNL VMM3a front-end ASIC and the RD51 Scalable Readout System (SRS).
    This allows to read out the detectors at high rates (up to the MHz regime) with electronics time resolutions of the order of $\SI{1}{ns}$ and the ability to handle different detector types and sizes, due to a larger dynamic range compared to the previous front-end electronics based on the APV25 ASIC.
    Having studied and improved the beam telescope's performance over the course of three test beam campaigns, the results are presented in this paper.}


\proceeding{7\textsuperscript{th} International Conference on Micro-Pattern Gaseous Detectors\\
  11--16 December 2022\\
  Rehovot, Israel}

\begin{document}
\maketitle
\flushbottom


\section{Introduction and system overview}
\label{sec:introduction}
As part of its activities, the RD51 collaboration~\cite{RD51} organises up to three test beam campaigns per year at the H4 extraction beam line of the CERN Super Proton Synchrotron (SPS).
For this, two beam telescopes are provided, one being based on COMPASS-like triple-GEM detectors~\cite{GEM_COMPASS} and the other on resistive-strip MicroMegas detectors~\cite{MM_RESISTIVE}.

Both telescopes are equipped with the common RD51 readout electronics---the Scalable Readout System (SRS)~\cite{SRS}---allowing to read out multiple detectors (several thousand readout channels) simultaneously.
Following the integration of the ATLAS/BNL VMM3a front-end ASIC~\cite{VMM3a} into the SRS~\cite{VMM3a_SRS_INTEGRATION, VMM3a_SRS_RATE}, it replaced the APV25~\cite{APV25} electronics of the GEM-based telescope.
This enables the acquisition of data at much higher rates (MHz regime instead of kHz), with the electronics providing charge and time of the induced signal simultaneously, with the latter being in the nanosecond scale.
In addition, different peaking times ($\num{25}$ to $\SI{200}{ns}$) and electronics gains ($\num{0.5}$ to $\SI{16}{mV/fC}$) can be selected, being useful in the R \& D context, as it allows tuning the electronics' settings to the specific detector types.
The VMM3a is operated in a self-triggered continuous readout mode, requiring precise clock synchronisation between the front-end cards.
This is achieved by using the SRS' Clock and Trigger Fanout (CTF) card.

The telescope contains three triple-GEM detectors ($\SI[product-units = power]{10x10}{cm}$ active area, $x$-$y$-strip anode, 256+256 strips at $\SI{400}{\mu m}$ pitch) over a distance of about $\SI{1}{m}$ to provide information on the beam particles' trajectories.
In addition, it contains three scintillators (two in front of the tracking detectors and one behind them) which are read out with Photo-Multiplier Tubes (PMTs).
The output pulses from the PMTs are processed by a NIM-logic including a coincidence unit, which is also read out with VMM3a/SRS.
This provides the arrival time of the interacting particles.

\section{Performance evaluation}

\subsection{Electronics' rate-capability}
\begin{figure}[t!]
    \centering
    \begin{subfigure}{0.49\textwidth}
        \centering
        \includegraphics[width = 0.75\textwidth]{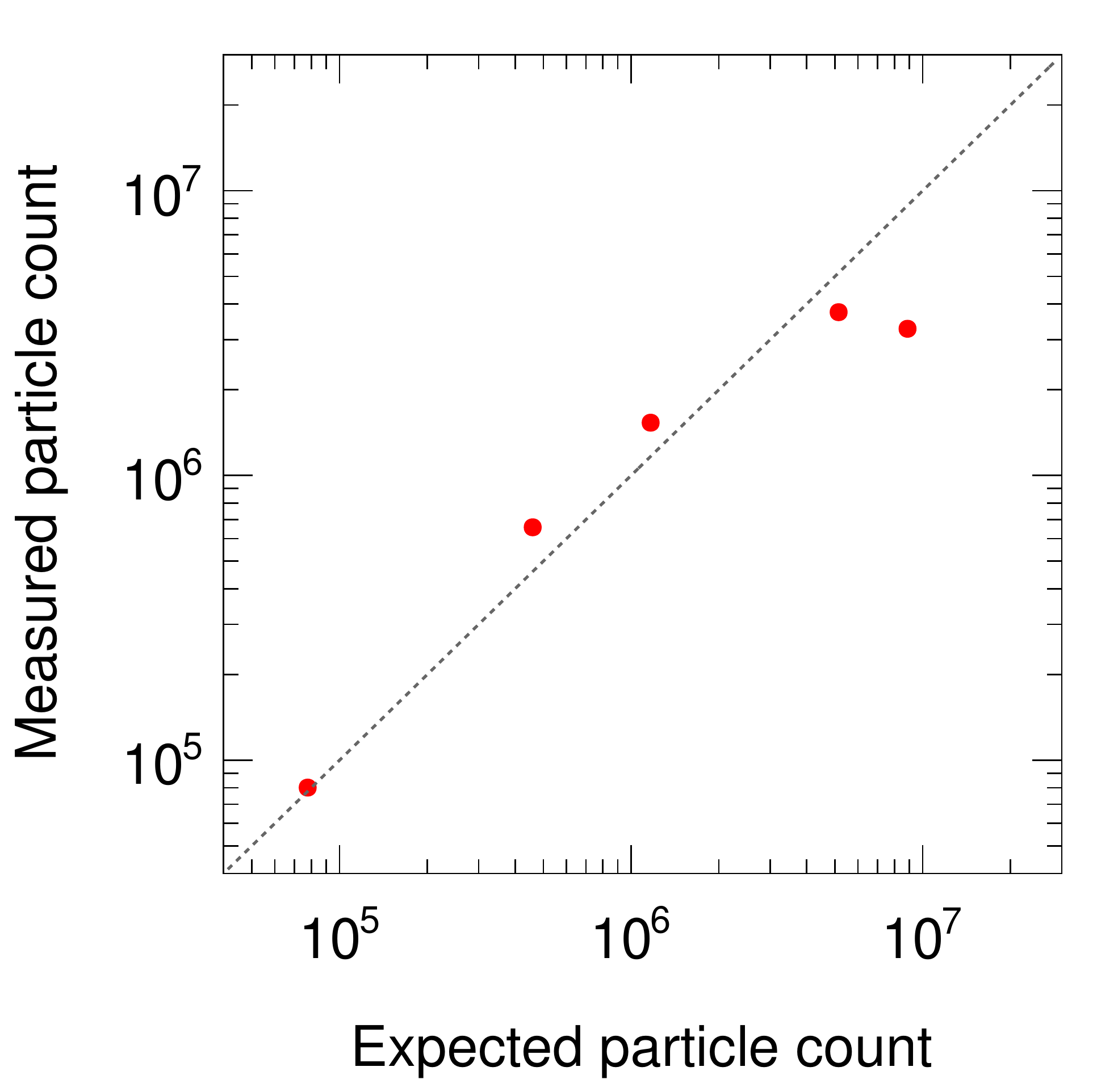}
        \caption{Rate-capability}
        \label{fig:rateCapabilityScan}
    \end{subfigure}
    \begin{subfigure}{0.49\textwidth}
        \centering
        \includegraphics[width = 0.75\textwidth]{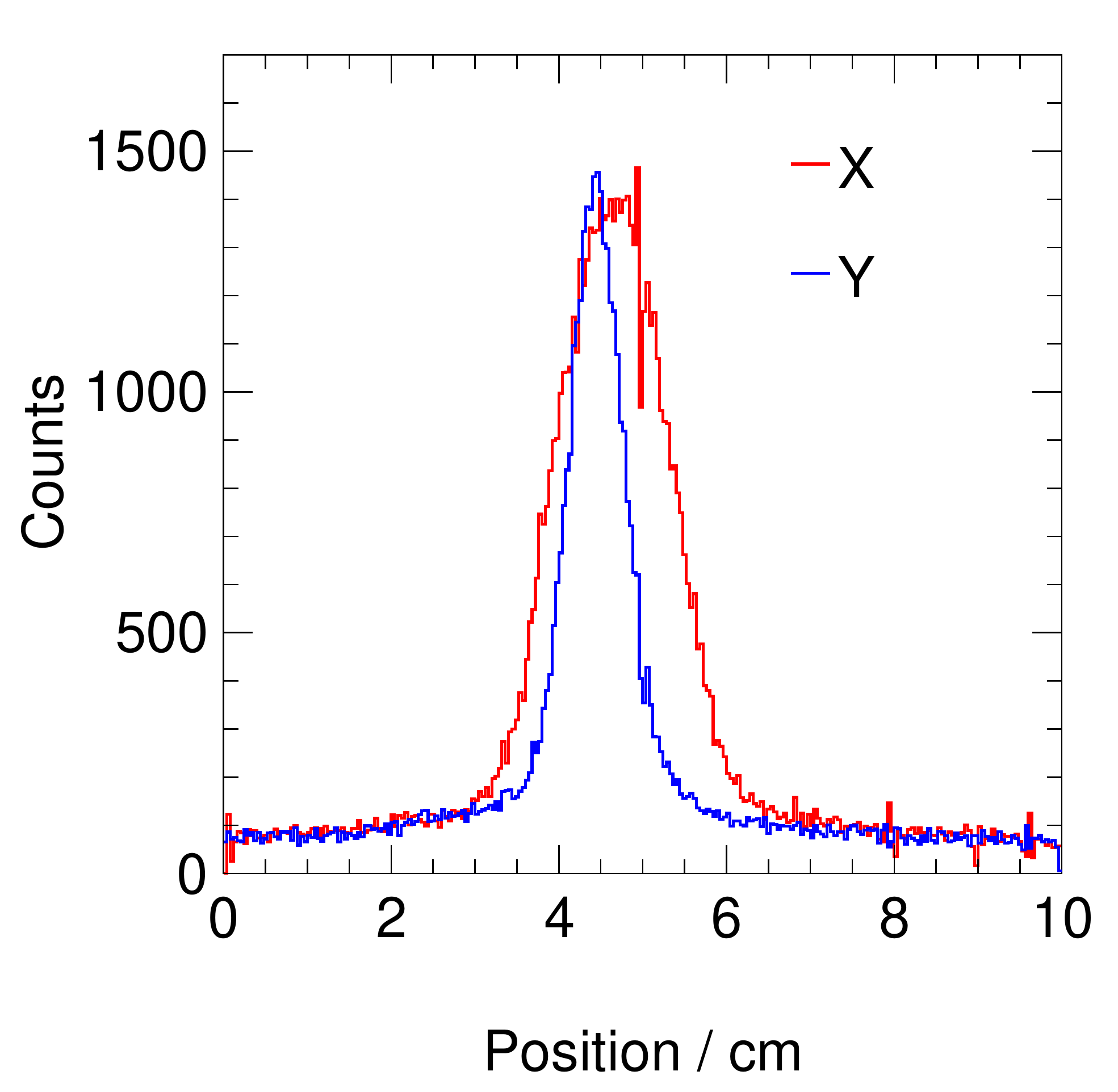}
        \caption{Beam profile}
        \label{fig:rateCapabilityProfile}
    \end{subfigure}
    \caption{Measured number of particles within a $\SI{5}{s}$ beam extraction versus the expected one (a).
        Transverse profile of an $\SI{80}{GeV/c}$ pion beam with $10^6$ particles (b).}
    \label{fig:rateCapability}
\end{figure}

VMM3a/SRS can acquire particle interaction rates in the MHz regime, following studies with X-rays~\cite{VMM3a_SRS_RATE}.
The same behaviour was observed with Minimum Ionising Particles (MIPs), where the number of particle interactions that can be recorded saturates at around $\SI{1}{MHz}$ (figure \ref{fig:rateCapabilityScan}).

This MHz rate-capability---in combination with the triple-GEM detectors---helps for example in understanding the beam structure.
While the expected number of particles is provided by the scintillators of the beamline instrumentation even at higher rates, the beam profile measurement of the beamline's delay wire chambers fails at rates $\gtrsim \SI{100}{kHz}$ (figure \ref{fig:rateCapabilityProfile}) due to the rate limits.

\subsection{Time resolution}
\label{sec:timeResolution}
\begin{figure}[t!]
    \centering
    \begin{subfigure}{0.49\textwidth}
        \centering
        \includegraphics[width = 0.75\textwidth]{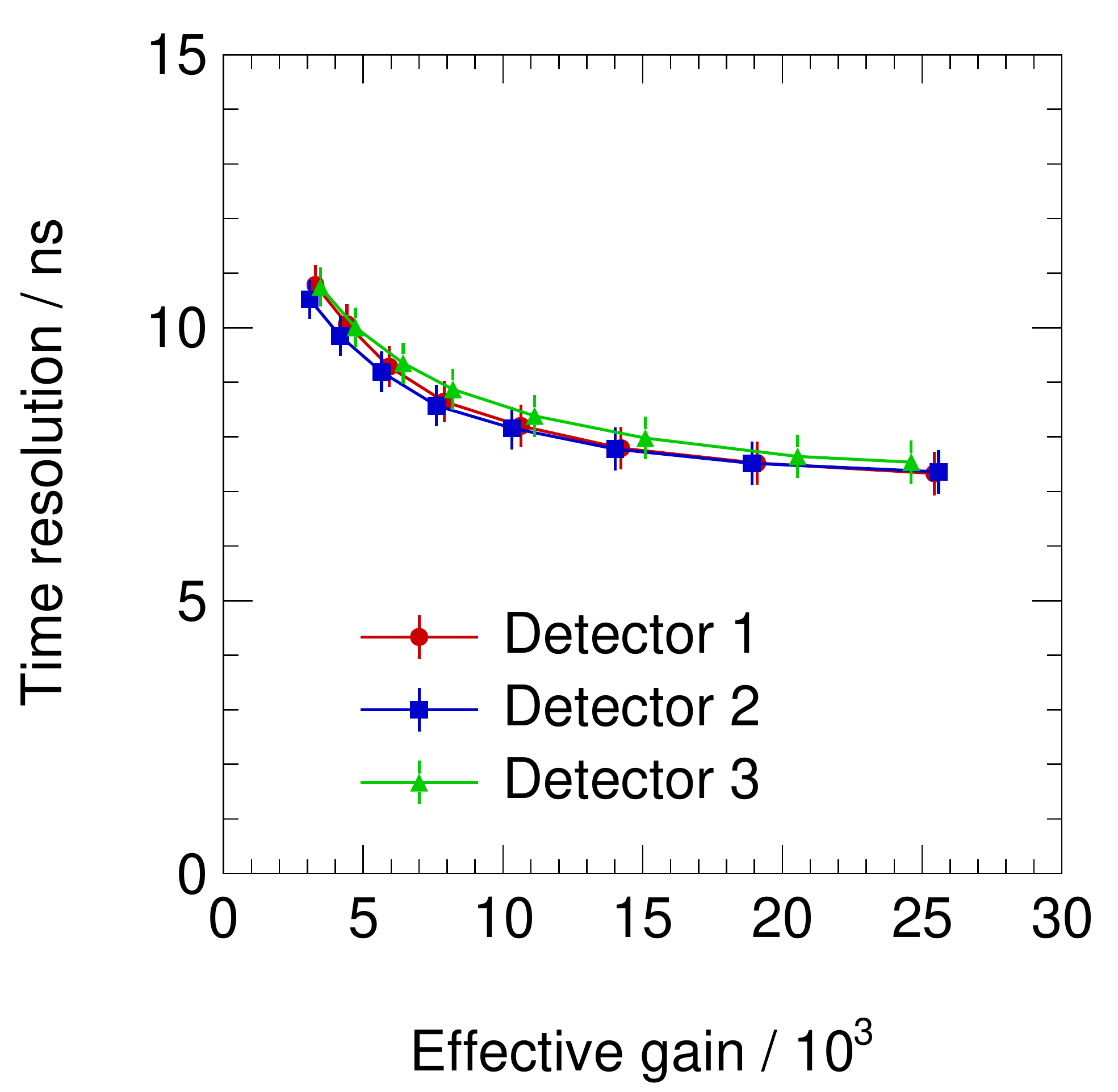}
        \caption{Tracking detectors}
        \label{fig:timeResolutionTrackers}
    \end{subfigure}
    \begin{subfigure}{0.49\textwidth}
        \centering
        \includegraphics[width = 0.9\textwidth]{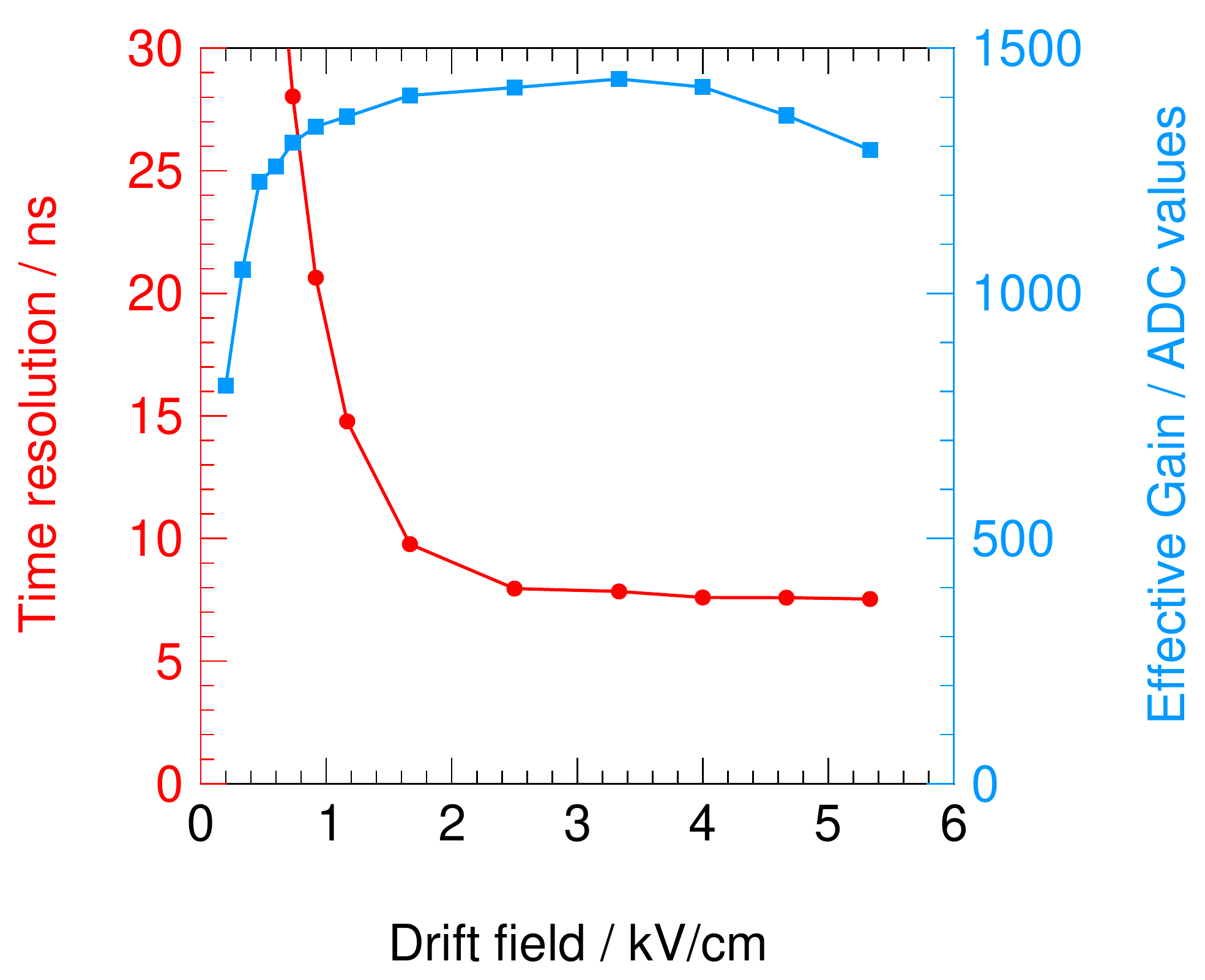}
        \caption{Detector under test}
        \label{fig:timeResolutionEnergyInformation}
    \end{subfigure}
    \caption{Time resolution of the three tracking detectors, depending on their gain (a).
        Time resolution of a COMPASS-like triple-GEM detector, depending on the drift field (b).}
    \label{fig:timeResolution}
\end{figure}

The VMM3a provides the signal arrival time with a resolution between $\num{0.5}$ and $\SI{2}{ns}$~\cite{VMM3a_SRS_TWEPP}.
This allows measuring the intrinsic time resolution of MPGDs for tracking without the need of external high-resolution electronics.
Using the pulses from the NIM coincidence unit as reference timestamps $t_\mathrm{ref}$, it is determined if a particle interaction was recorded by the detector of interest within a previously defined time window (e.g.\ $\pm \SI{500}{ns}$) around $t_\mathrm{ref}$.
The time of the particle interaction is defined as the arrival time of the signal with the largest amplitude $t_\mathrm{LA}$~\cite{VMMSDAT}, as the charge signal is typically spread over several adjacent readout channels.
For recorded interactions within the time window, $\Delta t = t_\mathrm{ref} - t_\mathrm{LA}$ is calculated, resulting in a Gaussian distribution.
From the width $\sigma_{\Delta t}$ of this distribution, the detector resolution can be extracted by quadratically subtracting the contributions from the front-end electronics, the NIM-logic and the scintillator/PMTs (figure \ref{fig:timeResolution}).

\subsection{Energy/charge information}
The matching in time between the coincidence unit's signals and the tracking detectors allows also for studying the charge/gain/energy behaviour of the detectors.
In the given example (figure \ref{fig:timeResolutionEnergyInformation}), the time resolution and charge collection behaviour of a triple-GEM detector are shown.

\subsection{Particle trajectories: spatial resolution and efficiency}
In the following, the spatial information is added, making use of the three tracking detectors and the reconstruction of the particles' trajectories with a Kalman filter.
Due to the self-triggered readout, the events for the track reconstruction are built during the offline analysis, using the recorded interactions in each detector
and a time window, similar to the time resolution studies (section \ref{sec:timeResolution}).

\begin{figure}[t!]
    \centering
    \begin{subfigure}{0.49\textwidth}
        \centering
        \includegraphics[width = 0.75\textwidth]{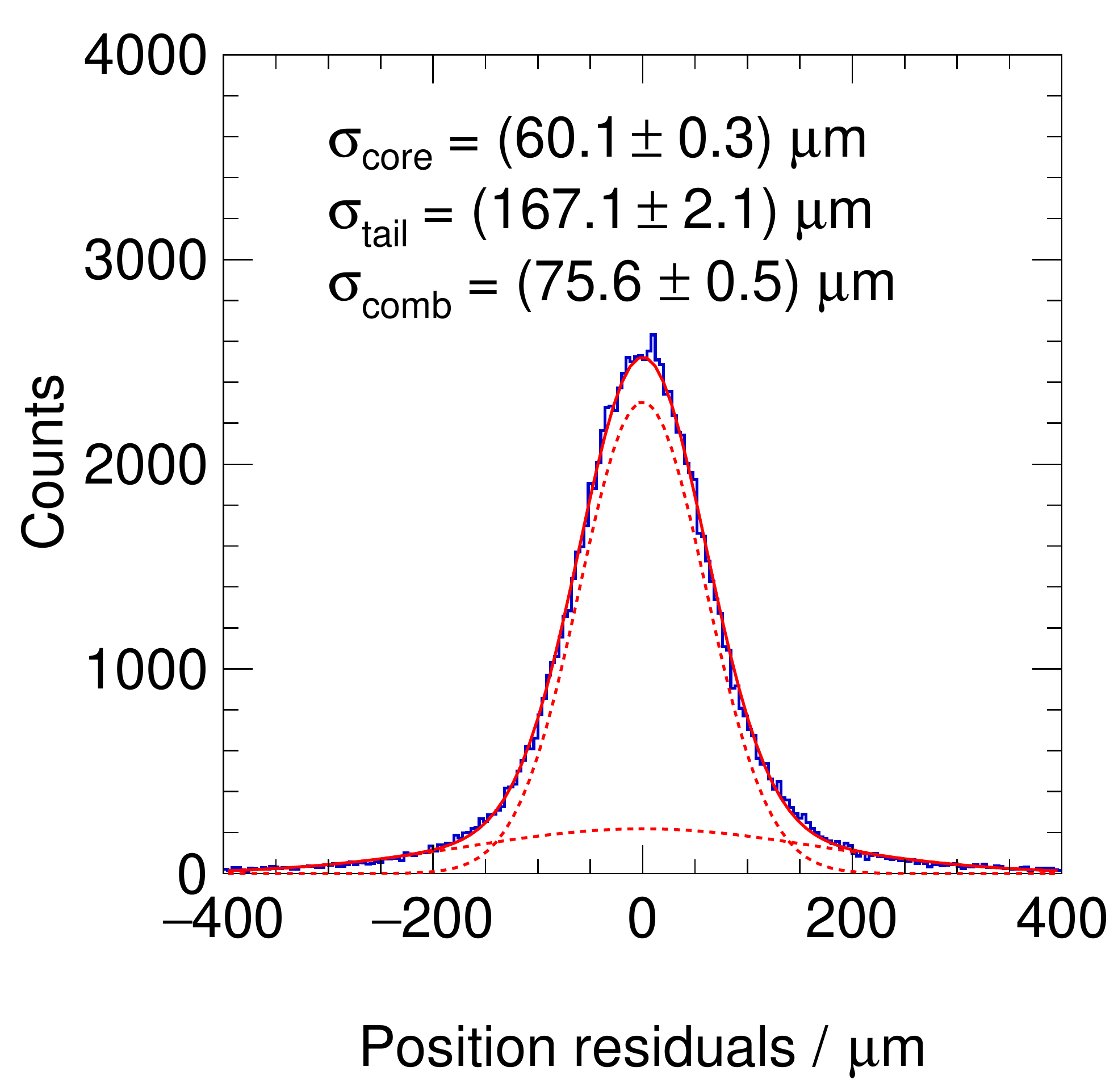}
        \caption{Residual distribution}
        \label{fig:spatialResolutionResiduals}
    \end{subfigure}
    \begin{subfigure}{0.49\textwidth}
        \centering
        \includegraphics[width = 0.9\textwidth]{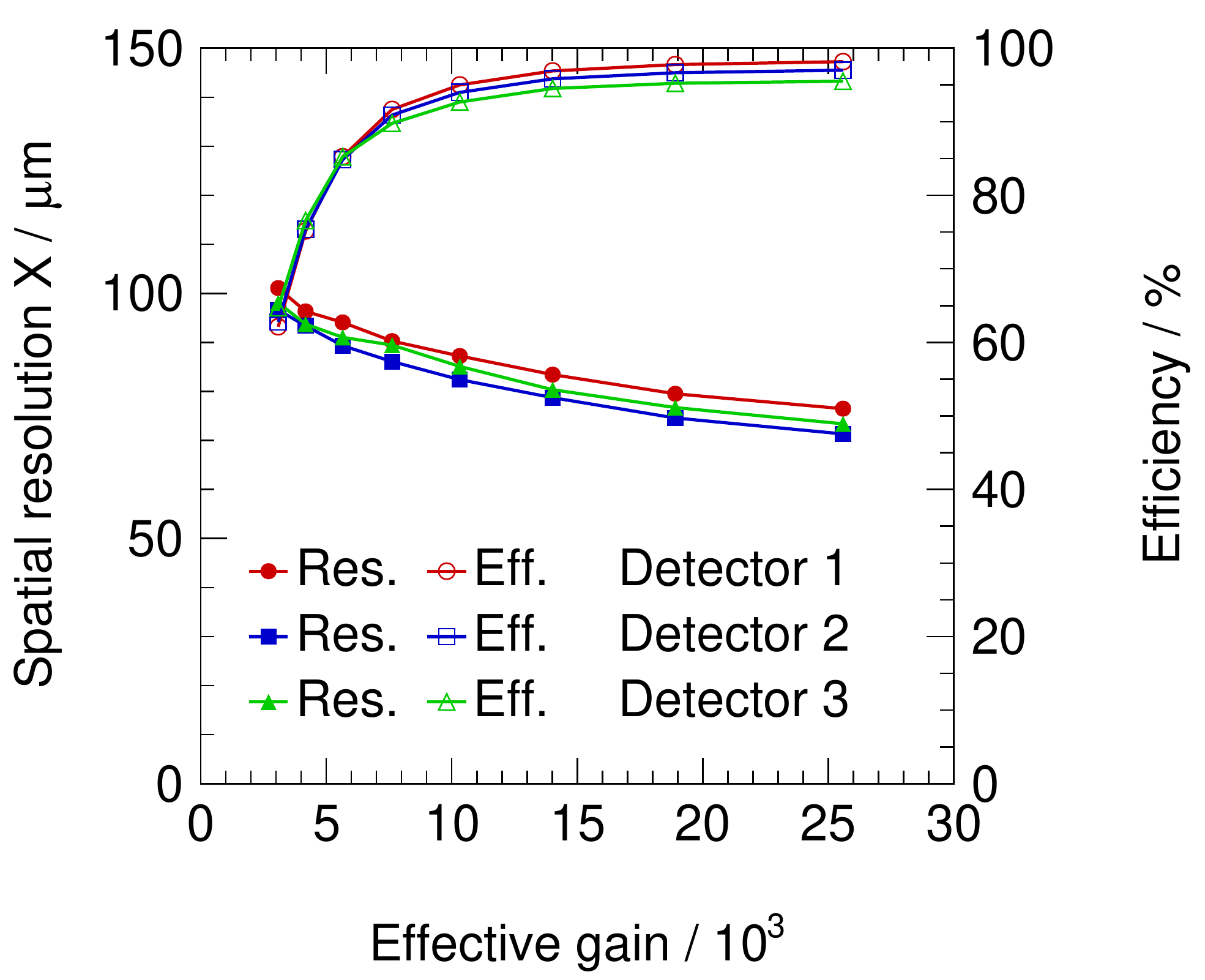}
        \caption{Spatial resolution (Res.) and efficiency (Eff.)}
        \label{fig:spatialResolutionTrackers}
    \end{subfigure}
    \caption{Example of the $\Delta x$ distribution (a).
        Spatial resolution and detector efficiency of the three tracking detectors (b).}
    \label{fig:spatialResolution}
\end{figure}

To determine the spatial resolution of the tracking detectors, the expected interaction point in the detector of interest $x_\mathrm{ref}$ is compared with the reconstructed one $x'$, leading to the distribution of position residuals $\Delta x = x_\mathrm{ref} - x'$ (figure \ref{fig:spatialResolutionResiduals}).
The distribution is fitted with a double-Gaussian function to account for the tails of the residual distribution.
The combined width of the residual distribution is determined by $\sigma_{\Delta x}^2 = w\sigma_\mathrm{centre}^2 + (1 - w)\sigma_\mathrm{tail}^2$, with the weighting factor $w$ being the relative scale between the centre and the tail distribution.
The detector resolution (figure \ref{fig:spatialResolutionTrackers}) is determined by subtracting quadratically the uncertainty of the track reconstruction.

The efficiency of the telescope's detectors is determined via $\epsilon_\mathrm{det} = N_\mathrm{comb} / N_\mathrm{ref}$, where $N_\mathrm{ref}$ is the number of expected particles (provided by the reference detectors) and $N_\mathrm{comb}$ is the number of recorded particle interactions where both, the reference detectors and the detector of interest, participate in the track fit.
Only at effective gains $>10^4$, efficiencies of $>\SI{90}{\%}$ can be reached.
The reason for this is the electronics threshold, which in this case was set to $\SI{1.5}{fC}$.
Due to the system being self-triggered at the threshold level (THL), each signal surpassing the THL is going to be processed, taking part of the available bandwidth.
As such, lowering the THL will lead at a certain point to a situation with electronic noise dominating the number of recorded signals.
A potential method to overcome the issue with noise acquisition at low threshold levels---still requiring further investigation---is the sub-hysteresis discrimination of the VMM (for details see~\cite{VMM1}).

\section{Improving the spatial resolution}
The telescope provides a precise reference position, making it the ultimate tool for studying improvements in spatial resolution; with a discrete readout structure and threshold-based electronics, charge information is lost, which decreases the accuracy of the position reconstruction when using the centre-of-gravity (COG) method (for details it is referred to~\cite{VMM3a_SRS_IMAGING}).
Two approaches---tested previously with X-rays~\cite{VMM3a_SRS_IMAGING}---are the VMM3a's neighbouring-logic\footnote{When the neighbouring-logic gets enabled and the signal on one channel surpasses the THL, the two adjacent channels are read out as well, even if the signal does not surpass the threshold level.} (hardware approach) and changing the $Q^2$ weighting method\footnote{In the centre-of-gravity formula, the position is calculated via $x = \sum_i Q_i^n x_i / \sum_i Q_i^n$ with $n = 1$. For the $Q^2$ weighting, $n = 2$ is used in order to reduce the weight of the tails of the charge distribution (for more details it is referred to~\cite{VMM3a_SRS_IMAGING}).} (software approach).

\begin{figure}[t!]
    \centering
    \begin{subfigure}{0.49\textwidth}
        \centering
        \includegraphics[width = 0.75\textwidth]{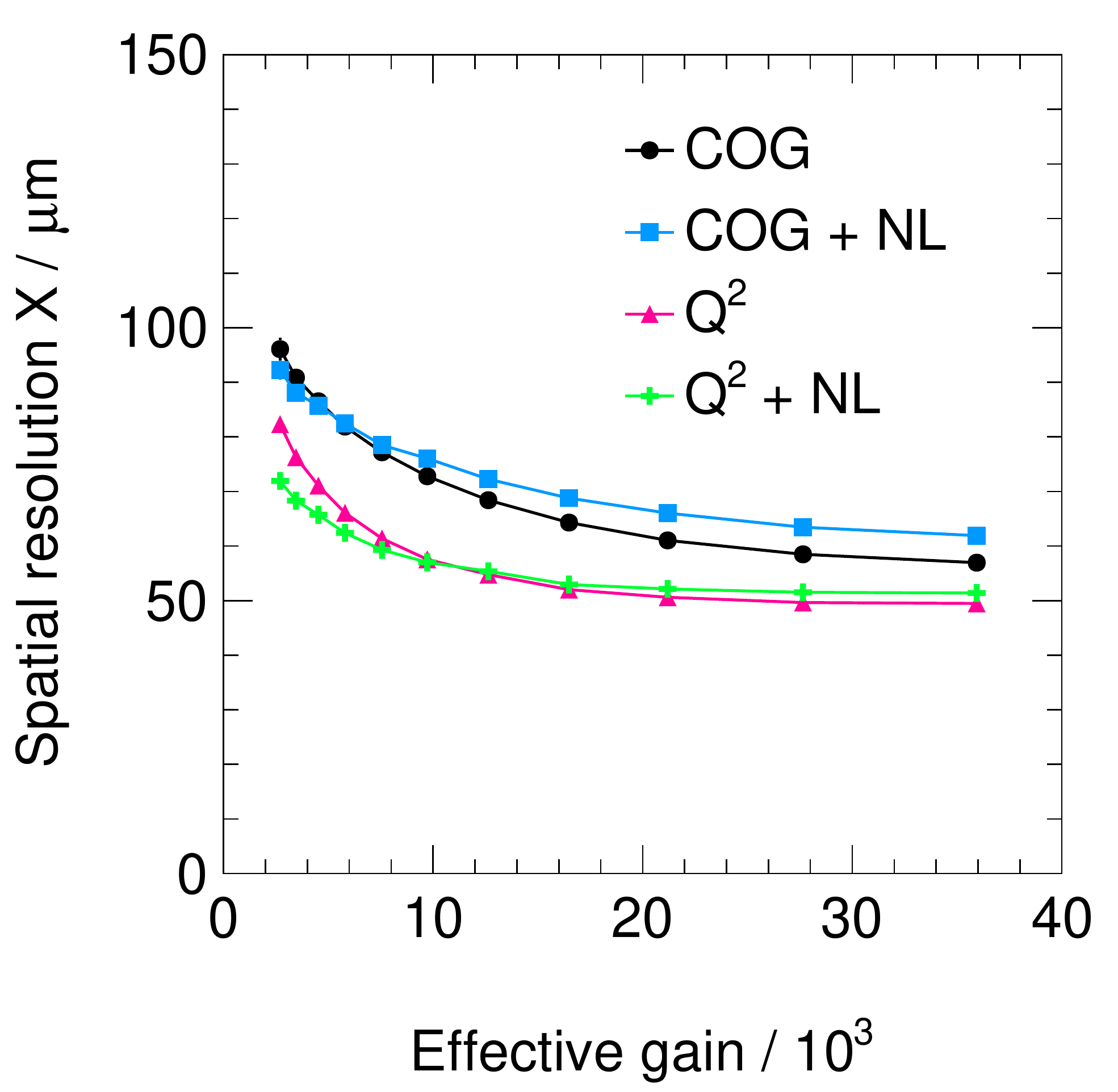}
        \caption{Spatial resolution}
        \label{fig:improvingResolutionMethods}
    \end{subfigure}
    \begin{subfigure}{0.49\textwidth}
        \centering
        \includegraphics[width = 0.75\textwidth]{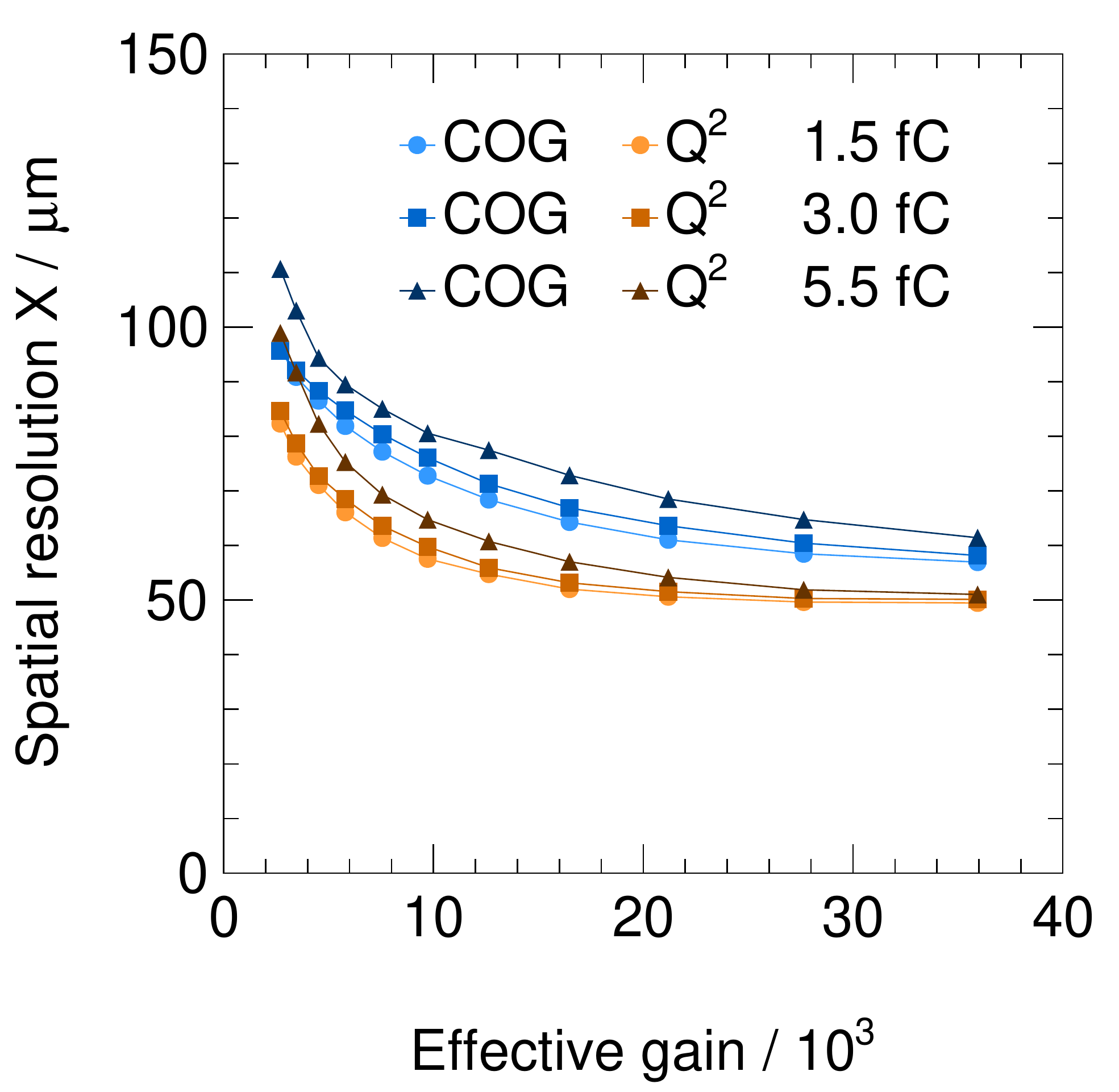}
        \caption{COG vs $Q^2$ weighting}
        \label{fig:improvingResolutionQ2Weighting}
    \end{subfigure}
    \caption{Spatial resolution at an electronics THL of $\SI{1.5}{fC}$ (a).
        Comparison between COG and $Q^2$ reconstruction at different THLs (b).}
    \label{fig:improvingResolution}
\end{figure}

A comparison between the two approaches and the COG results is shown in figure \ref{fig:improvingResolutionMethods}.
The neighbouring-logic improves the spatial resolution only in the low signal-to-threshold regime, i.e. at low detector efficiencies.
In these situations, the relative amount of charge recovered from below the THL is large, while at higher gains it is significantly lower, also with an increased probability of potentially acquiring noise which would decrease the spatial resolution.
The $Q^2$ weighting on the other hand improves the resolution over the full gain range, also for different threshold levels (figure \ref{fig:improvingResolutionQ2Weighting}), allowing to reach resolutions around $\SI{50}{\mu m}$.

\section{Conclusion}
With the integration of the VMM3a front-end ASIC into the RD51 Scalable Readout System, new possibilities opened for detector characterisation studies.
The commissioning of a beam telescope with VMM3a/SRS electronics allows studying detectors' time resolution (better than $\SI{10}{ns}$) and position resolution (around $\SI{50}{\mu m}$) simultaneously with a single readout system and all data being contained intrinsically in a single data stream.
At the same time, due to the increased rate-capability up to the MHz regime, the detector characterisation studies can be performed much faster, allowing to scan a larger range of parameters within a single test beam.
It should be noted that while the electronics' commissioning was performed with COMPASS-like triple-GEM detectors, other detector types ($\mu$RWELL~\cite{URWELL}, resistive-plane MicroMegas, triple-GEM detector with three-layer readout~\cite{XYU} or straw tubes~\cite{STRAWS}) have been also successfully studied with the new beam telescope and its electronics.

\acknowledgments

This work has been sponsored by the Wolfgang Gentner Programme of the German Federal Ministry of Education and Research (grant no.\ 13E18CHA).
This project has received funding from the European Union's Horizon 2020 Research and Innovation programme under Grant Agreement No 101004761.
The work has been supported by the CERN Strategic Programme on Technologies for Future Experiments (\url{https://ep-rnd.web.cern.ch/}).
The authors would like to thank Alexandru Rusu (CERN, SRS Technology) for the development and production of the readout electronics.


\end{document}